\documentclass{article}
\usepackage{ijcai19}

\usepackage{times}
\usepackage{soul}
\usepackage{url}
\usepackage[hidelinks]{hyperref}
\usepackage[utf8]{inputenc}
\usepackage[small]{caption}
\usepackage{graphicx}
\usepackage{amsmath}
\usepackage{booktabs}
\usepackage{algorithm}
\usepackage{algorithmic}
\usepackage{helvet}  
\usepackage{courier}  
\usepackage{graphicx}  
\usepackage{xcolor}
\usepackage{soul}
\usepackage{epsfig,amsfonts,amssymb,times,subfigure,floatflt,wrapfig,enumerate,tikz,url,bm}
\usepackage{siunitx}
\usepackage{tabu}
\usepackage{multirow}
\newtheorem{example}{Example}

\newtheorem{definition}{Definition}

\newcommand{\mat}[1]{\bm{#1}}
\newcommand{\ten}[1]{\bm{\mathcal{#1}}}

\allowdisplaybreaks[4]

\title{MiSC: Mixed Strategies Crowdsourcing}

\author{
Ching-Yun Ko$^1$
\and
Rui Lin$^1$\and
Shu Li$^2$\And
Ngai Wong$^1$
\affiliations
$^1$The University of Hong Kong, Hong Kong\\
$^2$Nanjing University, 210023, Nanjing, China
\emails
\{cyko, linrui, nwong\}@eee.hku.hk,
lis@smail.nju.edu.cn
}

\begin{document}

\maketitle

\begin{abstract}
Popular crowdsourcing techniques mostly focus on evaluating workers' labeling quality before adjusting their weights during label aggregation. Recently, another cohort of models regard crowdsourced annotations as incomplete tensors and recover unfilled labels by tensor completion. However, mixed strategies of the two methodologies have never been comprehensively investigated, leaving them as rather independent approaches. In this work, we propose \textit{MiSC} (\textbf{Mi}xed \textbf{S}trategies \textbf{C}rowdsourcing), a versatile framework integrating arbitrary conventional crowdsourcing and tensor completion techniques.
In particular, we propose a novel iterative Tucker label aggregation algorithm that outperforms state-of-the-art methods in extensive experiments. 
\end{abstract}

\section{Introduction}
 In recent years, with the advent of many complex machine learning models, the need for large amounts of labeled data is boosted. Though we can obtain unlabeled data abundantly and cheaply, acquiring labeled data from domain experts or well-trained workers with specific background knowledge is usually expensive and time-consuming. Crowdsourcing provides an effective way of collecting labeled data quickly and inexpensively. However, as crowd workers are usually non-experts or even spammers, the individual contribution from one worker can be unreliable. To improve the label quality, each item is presented to multiple workers. Therefore, inferring the true labels from a large sum of noisy labels is a key challenge in crowdsourcing.

Existing learning-from-crowds works mainly focus on eliminating annotations~\cite{li2018Multi} from unreliable workers or reducing their weights. In line with this goal, a series of work focusing on evaluating workers' quality are proposed based on accuracy~\cite{whitehill2009whose,karger2011iterative}, and confusion matrix~\cite{dawid1979maximum,raykar2009supervised,liu2012variational,zhou2012learning,zhang2016spectral}.
Apart from that, two emerged works~\cite{zhou2016crowdsourcing} and~\cite{li2018Multi} aim at completing annotations by translating single/multi-label crowdsouring tasks to tensor completion problems. However, their improvements in performance over conventional annotations methodologies are limited. 
In this work, we present a label aggregation algorithm by mixing the two strategies: We view the label tensor as an incomplete and noisy tensor, which is coherent to the reality where incompleteness comes from heavy workloads and noise stems from mislabelling. 
In a nutshell, we capture the structural information and filter out noisy information from the label tensor through tensor completion. This is followed by a conventional label aggregation procedure. We iterate over the above two steps until convergence. Despite the generality of the proposed framework, we highlight our use of tensor decomposition techniques.

Tensors are a higher-order generalization of vectors and matrices and constitute a natural representation for many real-life data that are intrinsically multi-way. In analogy to the significance of matrix QR factorization and singular value decomposition in matrix preconditioning and principal component analysis, tensor decomposition concepts have been deployed in modern machine learning models. Iconic examples include text analysis~\cite{collins2012tensor,liu2015learning}, compression of neural networks~\cite{denton2014exploiting,novikov2015tensorizing}, and tensor completion~\cite{suzuki2015convergence,imaizumi2017tensor}. Among various popular tensor decomposition techniques, we specifically have interests in the Tucker model~\cite{tucker1963implications,levin1963three,de2000multilinear}. The Tucker decomposition factorizes a tensor into a core tensor of generally smaller size, along with factor matrices for each of the tensor modes. We exemplify through a case study and show that for a perfectly labeled matrix, its corresponding label tensor will have an intrinsically low-rank Tucker decomposition. The main contributions of this article are:
\begin{itemize}
\item A general mixed strategies framework for crowdsourcing tasks is proposed, in which annotations are completed by tensor recovery and ground-truth labels are estimated by conventional deduction algorithms.
\item To our knowledge, this is the \emph{first work} that introduces tensor decomposition methods to exploit the structural information in the label tensor. This scheme also bears a clear physical interpretation.
\item Experimental results demonstrate that the proposed algorithms manage to improve existing methods to achieve higher aggregation accuracy. 
\item Most importantly, the proposed algorithms are shown to be particularly powerful compared to conventional pure label aggregation methods when the annotations are highly sparse and severely noisy. This is crucial since obtaining low-sparsity/ high-quality annotations can be arduous and expensive.
\end{itemize}
In the following, Section~\ref{sec:related} briefly summarizes the related work. Then, Section~\ref{sec:preliminaries} presents the necessary tensor preliminaries. Section~\ref{sec:mod} focuses on one of the proposed models, mixed low-rank Tucker DS-EM aggregation, and showcases the algorithm. Next, numerical experiments comparing the proposed MiSC (mixed strategies crowdsourcing) with pure label aggregation methods are given in Section~\ref{sec:experiments}. Finally, Section~\ref{sec:conclusion} concludes this work.

\section{Related Work}
\label{sec:related}
Our two-stage mixed algorithms build upon the standard label aggregation algorithms. Majority voting, as the most straightforward crowdsourcing technique, has served as a baseline method for years. Besides, the seminal work of Dawid and Skene based on expectation maximization (EM), denoted as DS-EM henceforth, is also among the earliest work of crowdsourcing~\cite{dawid1979maximum} which is a generative method by assuming the performance of each worker is consistent across different tasks. 
Moreover, the crowdsourcing problem is further translated into a variational Bayesian inference problem of a graphical model in~\cite{liu2012variational}. Using Mean Field algorithm, the parameters are tuned to maximize the marginal likelihood.
~\cite{zhou2012learning} introduces the minimax entropy principle into crowdsourcing tasks and infers true labels by minimizing the Kullback–Leibler (KL) divergence between the probability distribution over workers, items, labels, and the ground truth. 
 
In matrix completion, most problems are formulated into the construction of a structurally low-rank matrix $\mat{X}$ having the same observed entries: 
$\min_{\mat{X}} {\textrm{rank}(\mat{X})}, \textrm{ \textrm{s.t.} } (\mat{X}-\mat{A})_{\Omega} = \mat{0}$, where $\mat{A}$ represents the matrix with missing values filled by zeros and $\Omega$ is the mapping that specifies the locations of originally non-zero elements. Directly solving the above optimization problem is NP-hard, which results in extensive research on solving alternative formulations. One of the two popular candidates is to minimize the nuclear norm as the convex envelope of the matrix rank-operator~\cite{candes2009exact,chen2015incoherence}. This nuclear norm minimization idea is then further generalized to tensor completion problems by computing matricizations of the tensor along its $k$ modes and summing over the nuclear norm of the resulting $k$ matrices (abbreviated as LRTC in the remainder of this paper)~\cite{liu2013tensor,signoretto2014learning}.
Methods that exploit tensor decomposition formats were also introduced to tensor completion problems in recent years. In~\cite{Prateek2014Provable,suzuki2015convergence,zhao2015bayesian}, the authors use the Canonical Polyadic (CP) decomposition for tensor estimators. 

The pioneers of introducing tensor completion concepts to crowdsourcing problems are~\cite{zhou2016crowdsourcing} and~\cite{li2018Multi}, where authors build label aggregation algorithms upon the methodologies introduced in~\cite{liu2013tensor}. This work contrasts with them in two main aspects: (1) \textit{Aims} - we focus on introducing a versatile complete-aggregate two-step looping structure for crowdsourcing tasks, where completion techniques adopted can be of any kind. (2) \textit{Approaches} - we introduce tensor decomposition driven completion algorithms and showcase its advantages over LRTC methods in crowdsourcing tasks.

\section{Preliminaries}
\label{sec:preliminaries}
Tensors are high-dimensional arrays that generalize vectors and matrices. In this paper, boldface capital calligraphic letters $\ten{A},\ten{B},\ldots$ are used to denote tensors, boldface capital letters $\mat{A},\mat{B},\ldots$ denote matrices, boldface letters $\mat{a},\mat{b},\ldots$ denote vectors, and Roman letters $a,b,\ldots$ denote scalars. A $d$-way or $d$-order tensor \mbox{$\ten{A} \in\mathbb{R}^{I_1\times I_2\times \cdots\times I_d}$} is an array where each entry is indexed by $d$ indices $i_1,i_2,\ldots,i_d$. We use the convention $1\leq i_k\leq I_k$ for $k=1,\ldots,d$, where the $I_k$ is called the $k$th dimension of tensor $\ten{A}$. MATLAB notation $\ten{A}(i_1,i_2,\ldots,i_d)$ is used to denote entries of tensors. Fibers are high-dimensional analogue of rows and columns in matrices. In a matrix $\mat{A}$, a matrix column can be referred by fixing a column index. By again adopting MATLAB convention, the $i_2$th column of matrix $\mat{A}$ is denoted $\mat{A}(:,i_2)$, which is also called a \mbox{$1$-mode} fiber of the matrix. Similarly, a matrix row $\mat{A}(i_1,:)$ is known as a \mbox{$2$-mode} fiber of the matrix. For a $d$-way tensor, a \mbox{$k$-mode} fiber is determined by fixing all the indices but the $k$th one, which we denote as \mbox{$\ten{A}(i_1,\ldots,i_{k-1},:,i_{k+1},\ldots,i_d)$}.

Tensor $k$-mode matricization, also known as tensor $k$-mode unfolding/flattening, reorders the elements of a $d$-way tensor into a matrix. It is formally defined as:
\begin{definition}\cite[p.~459]{tensorreview} 
The $k$-mode matricization of a tensor \mbox{$\ten{A}\in\mathbb{R}^{I_1\times\cdots\times I_d}$}, denoted by \mbox{$\mat{A}_{(k)}\in\mathbb{R}^{I_k\times I_1\cdots I_{k-1} I_{k+1}\cdots I_d}$}, arranges the $k$-mode fibers to be the columns of the resulting matrix. Tensor element \mbox{$\ten{A}(i_1,\ldots,i_d)$} maps to matrix element $\mat{A}_{(k)}(i_k,j)$, where
$j=1+\sum^{d}_{p=1,p\neq k}(i_p-1)J_p$ and $J_p=\prod^{p-1}_{m=1,m\neq k}I_m$.
\end{definition}
The generalization of the matrix-matrix multiplication to tensor involves a multiplication of a matrix with a $d$-way tensor along one of its $d$ modes:
\begin{definition}\cite[p.~460]{tensorreview}
The $k$-mode product of a tensor~\mbox{$\ten{G}\in\mathbb{R}^{R_1\times\cdots \times R_d}$} with a matrix $\bm{U}\in\mathbb{R}^{J \times R_k}$ is denoted \mbox{$\ten{A}=\ten{G}\, {\times_k}\, \bm{U}$} and defined by%
\begin{align*}
\small \ten{A}(r_1,\cdots,r_{k-1},j,r_{k+1}, \cdots, r_d) &= \hfill \\
\small\sum_{r_k=1}^{R_k}  \mat{U}(j,r_k) \ten{G}(r_1,\cdots,r_{k-1},r_k,&r_{k+1},\cdots,r_d),
\end{align*}
where $\ten{A}\in\mathbb{R}^{R_1\times\cdots \times R_{k-1}\times J \times R_{k+1}\times\cdots\times R_d}$.\label{def:kmode}
\end{definition}
Following Definition~\ref{def:kmode}, the full multilinear product of a $d$-way tensor with $d$ matrices is:
\begin{definition}\cite[p.~147]{cichocki2015tensor}
The full multilinear product of a tensor~\mbox{$\ten{G}\in\mathbb{R}^{R_1\times\cdots \times R_d}$} with matrices $\bm{U}^{(1)}, \bm{U}^{(2)},\ldots,\bm{U}^{(d)}$, where $\bm{U}^{(k)}\in\mathbb{R}^{I_k \times R_k}$, is denoted $\ten{A}=[[\ten{G}; \bm{U}^{(1)}, \bm{U}^{(2)},\ldots,\bm{U}^{(d)}]]$ and defined by $\ten{A} = \ten{G}\, {\times_1}\, \bm{U}^{(1)} \, {\times_2}\, \bm{U}^{(2)} \cdots {\times_d}\, \bm{U}^{(d)}$, where \mbox{$\ten{A}\in\mathbb{R}^{I_1\times\cdots\times I_d}$}.
\end{definition}
Following the definition of the full multilinear product, we define the Tucker decomposition as follows:
\begin{definition}
The Tucker decomposition decomposes a tensor \mbox{$\ten{A}\in\mathbb{R}^{I_1\times\cdots\times I_d}$} into a core tensor \mbox{$\ten{G}\in\mathbb{R}^{R_1\times\cdots\times R_d}$} multiplied by a matrix $\mat{U}^{(k)}\in\mathbb{R}^{I_k \times R_k}$ along the $k$th mode for $k=1,\ldots,d$. That is, a \mbox{$d$-way} tensor $\ten{A}$ is regarded as a multilinear transformation of a small core tensor $\ten{G}$ by $d$ factor matrices $\mat{U}^{(1)},\mat{U}^{(2)},\ldots,\mat{U}^{(d)}$. By writing out $\mat{U}^{(k)}=[\mat{u}_1^{(k)},\mat{u}_2^{(k)},\ldots,\mat{u}_{R_k}^{(k)}]$ for $k=1,2,\dots,d$, we have
\vspace{-0.45em}
\begin{align}
\label{eqn:tucker1}
\small\ten{A}&=\small\sum\limits^{R_1}_{r_1=1}\cdots\sum\limits^{R_d}_{r_d=1}\ten{G}(r_1,\ldots,r_d)(\mat{u}_{r_1}^{(1)}\circ\cdots\circ\mat{u}_{r_d}^{(d)}),\\
\small\nonumber&= \small\ten{G}\, {\times_1}\, \bm{U}^{(1)} \, {\times_2}\, \bm{U}^{(2)} \cdots {\times_d}\, \bm{U}^{(d)},\\
\small\nonumber&=\small[[\ten{G}; \bm{U}^{(1)}, \bm{U}^{(2)},\ldots,\bm{U}^{(d)}]],
\end{align}
where $r_1,r_2,\ldots,r_{d}$ are auxiliary indices that are summed over, and $\circ$ denotes the outer product. 
\end{definition}
The dimensions $(R_1,R_2,\ldots,R_{d})$ of these auxiliary indices are called the Tucker ranks. It is worth noting that $R_k$ is in general no bigger than $rank(\mat{A}_{(k)})$, which is also called the multilinear rank. In other words, for a Tucker representation with Tucker ranks $(S_1,S_2,\ldots,S_{d})$, if there exists a $k$, where $1\leq k\leq d$ and $S_k> rank(\mat{A}_{(k)})$, then we can always find an equivalent Tucker representation with Tucker ranks no bigger than the multilinear ranks.

When the core tensor $\ten{S}$ is cubical and diagonal, a Tucker model reduces to a CP model. By writing out the formula, a CP decomposition expresses a $d$-way tensor $\ten{A}$ as 
\vspace{-0.4em}
\begin{align*}
\small\ten{A}=\sum\limits^{R}_{r=1}\ten{G}(r,\ldots,r)(\mat{u}_{r}^{(1)}\circ\cdots\circ\mat{u}_{r}^{(d)}).
\end{align*}

\begin{figure}[tb]
\begin{center}
\includegraphics[width=0.4\textwidth]{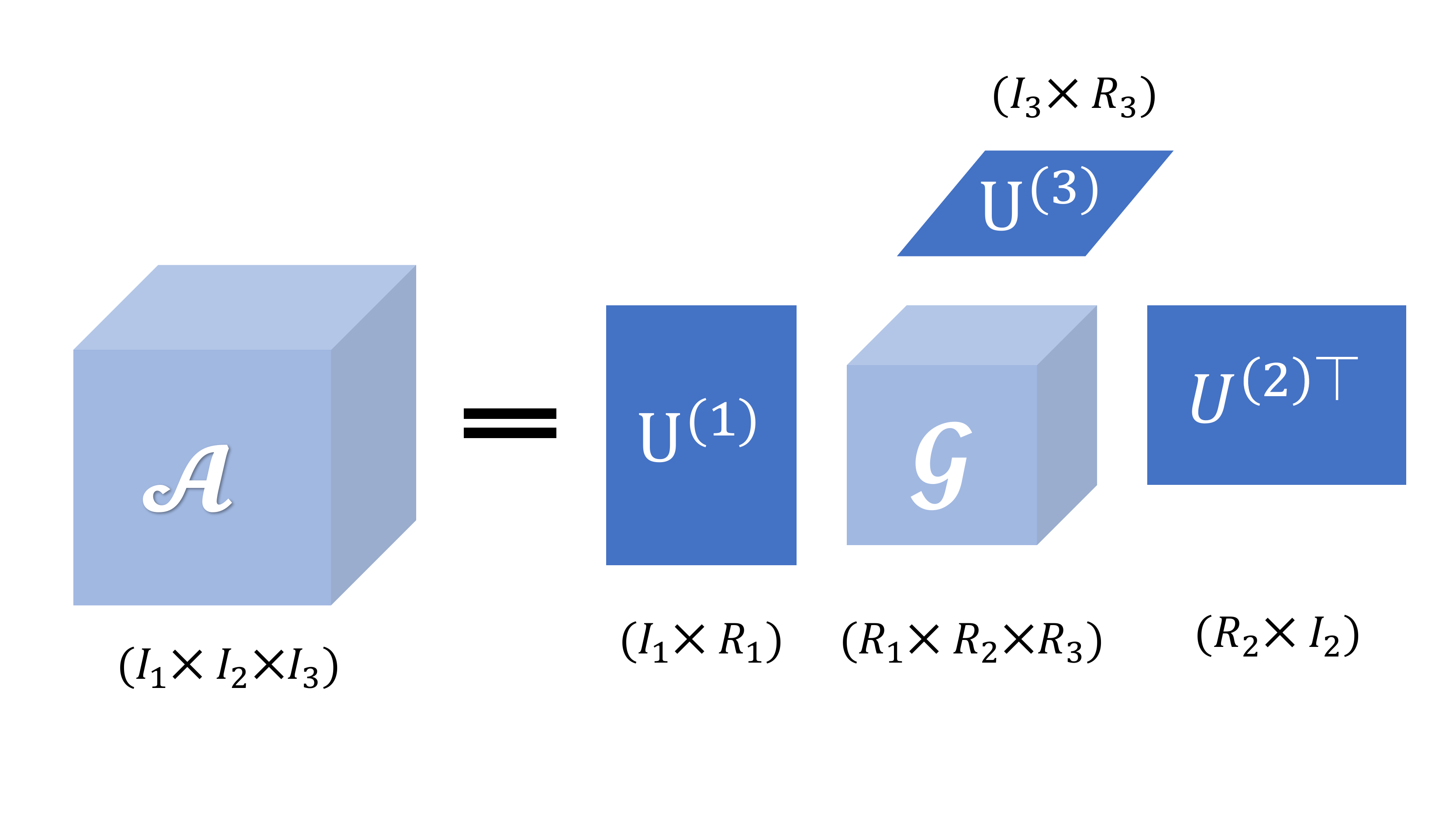}
\vspace{-1.5em}
\caption{Graphical depiction of the Tucker decomposition of a $3$-way tensor $\ten{A}$. $\ten{G}$ represents the core tensor and $\mat{U}^{(1)}, \mat{U}^{(2)},\mat{U}^{(3)}$ are the factor matrices.}
\label{fig:TN}
\end{center}
\vspace{-1em}
\end{figure}

\section{When Label Aggregation Meets Tensor Completion}
\label{sec:mod}
We now exemplify the proposed complete-aggregate two-step looping scheme using the case of mixed low-rank Tucker DS-EM aggregation. Firstly, we showcase in Section~\ref{subsec:basic} that Tucker model bears a clear physical meaning in completing label tensor. In Section~\ref{subsec:methodology}, we explain in detail how to combine a low-rank Tensor completion with traditional label aggregation to form one loop. The MiSC algorithms are then given in~\ref{subsec:algo}.
\subsection{Basic Idea}
\label{subsec:basic}
We use $\mat{A}\in\mathbb{R}^{N_w\times N_i}$ to denote the label matrix, where $N_w$ is the number of workers and $N_i$ is the number of items. If item $i$ is labeled by worker $w$ as class $c$, then we have $\mat{A}(w,i)=c$. Now we proposed to form a three-way binary tensor $\ten{A}$ of size $N_w\times N_i\times N_c$ by using the label matrix. This is done by using the canonical basis vectors to denote the non-zero elements in the matrix and all-zero vectors to denote unlabeled entries. Due to the fact that each worker gives at most one label for one item, there then contains at most one $1$ in each of the $3$-mode fibers of tensor $\ten{A}$. We use a small example to illustrate this process.

\begin{example}
Suppose there are two workers labeling for three items among four classes. The first worker believes the first item and the third item belong to the first class and the fourth class, respectively. The second worker labels the first item as the first class and the second item as the third class. We use label matrix $\mat{A}\in\mathbb{R}^{2\times 3}$ to describe the above setting,  where
\begin{align*}
\small\mat{A}=
\small\begin{pmatrix}
1 & 0 & 4\\
1 & 3 & 0\\
\end{pmatrix}.
\end{align*}
Considering matrix $\mat{A}$ contains four non-zero elements, four $3$-mode fibers of the resulting tensor contain a $1$, while the remaining two fibers are initiated with zeros vectors. For $\mat{A}(1,1)=1$ and $\mat{A}(2,1)=1$, the canonical basis vector $\mat{e}_1=(1,0,0,0)^T$ is used. For $\mat{A}(1,3)=4$ and $\mat{A}(2,2)=3$, the canonical basis vectors $\mat{e}_4=(0,0,0,1)^T$ and $\mat{e}_3=(0,0,1,0)^T$ are needed respectively. Collecting these fibers renders the following tensor 
\begin{align*}
\small\ten{A}(1,:,:)=
\small\begin{pmatrix}
1 & 0 & 0 & 0\\
0 & 0 & 0 & 0\\
0 & 0 & 0 & 1
\end{pmatrix},
\small\ten{A}(2,:,:)=
\small\begin{pmatrix}
1 & 0 & 0 & 0\\
0 & 0 & 1 & 0\\
0 & 0 & 0 & 0
\end{pmatrix}.
\end{align*}
\label{exm:exm1}
\end{example}
\vspace{-0.4em}
We call the above-defined tensor $\ten{A}$ label tensor. Our goal is to complete the label tensor $\ten{A}$ by assuming a specific underlying structure. Hence the problem can be regarded as a tensor completion task. As is well-known, the quality of the label matrix is in general not guaranteed. Therefore, it is of great importance that the proposed method should not only fill in entries but also serve as a pre-processor to de-noise the label tensor before doing deduction. 

Similar to the traditional matrix completion problem formulations, in this paper we propose  to form the following optimization problem:
\begin{align}
\label{eqn:optmization}
&\min_{\ten{G}, \bm{U}^{(1)},\bm{U}^{(2)},\bm{U}^{(3)}} \;||[[\ten{G}; \bm{U}^{(1)}, \bm{U}^{(2)},\bm{U}^{(3)}]] - \ten{A}||,\\
\label{eqn:constraint}
\textrm{\textrm{s.t.}} &\;\text{rank}_{\textrm{Tucker}}([[\ten{G}; \bm{U}^{(1)}, \bm{U}^{(2)},\bm{U}^{(3)}]]) = (R_1,R_2,R_3).
\end{align}
The optimization problem is solved via a truncated Tucker decomposition. It is worth noting that finding an exact Tucker decomposition of rank $(R_1,R_2,R_3)$, where $R_k=rank(\mat{A}_{(k)})$, of $\ten{A}$ is easy. However, finding an optimal truncated Tucker decomposition is nontrivial~\cite{tensorreview}. In this paper, the Tucker model is initialized using a truncated higher-order SVD~\cite{de2000multilinear} and updated by a higher-order orthogonal iteration algorithm~\cite{de2000best}. We now use an extreme case to demonstrate the vivid physical meaning behind the choice of a Tucker model.
\begin{example}
Suppose the label matrix is perfectly given by two workers for three items among four classes as follows
\begin{align*}
\small\mat{A}=
\small\begin{pmatrix}
1 & 3 & 4  \\
1 & 3 & 4  \\
\end{pmatrix}.
\end{align*}
The label tensor is then constructed as 
\begin{align*}
\small\ten{A}(1,:,:)=\ten{A}(2,:,:)=
\small\begin{pmatrix}
1 & 0 & 0 & 0\\
0 & 0 & 1 & 0\\
0 & 0 & 0 & 1\\
\end{pmatrix}.
\end{align*}
Now we consider the exact Tucker decomposition of this label tensor. By taking three different mode matricizations, we obtain
$rank(\ten{A}_{(1)})= 1, rank(\ten{A}_{(2)})= 3, rank(\ten{A}_{(3)})= 3$,
where $\min{(N_i,N_c)}=\min{(3,4)}=3$.
\label{exm:exm2}
\end{example}
As is demonstrated in the above example, if given a noise-less label tensor, it can be naturally decomposed into an exact low-rank Tucker format with Tucker ranks equal to $(1,\min{(N_i,N_c)},\min{(N_i,N_c)})$.

\subsection{Methodology}
\label{subsec:methodology}
\begin{figure}[t] 
\begin{center} 
\includegraphics[width=0.47\textwidth]{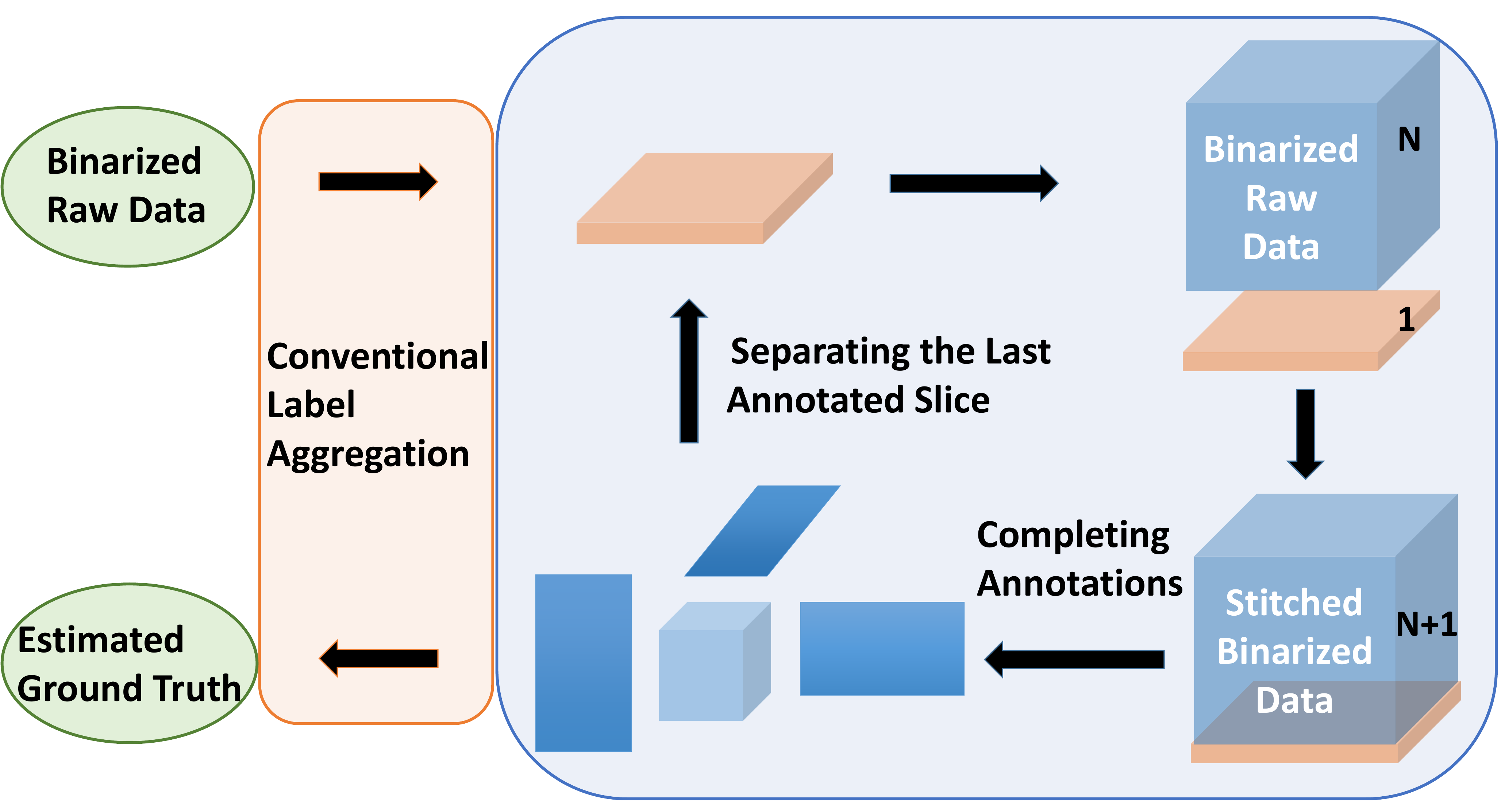}
\caption{The MiSC work flow.}
\label{fig:flow}
\end{center}
\vspace{-1.0em}
\end{figure}
In general, the proposed MiSC consists of two phases: label tensor completion and deduction. In each phase, a wide range of algorithms are available as will be seen in Section~\ref{sec:experiments}. We visualize the architecture of MiSC in Figure~\ref{fig:flow}.

The method starts from a binarized label tensor constructed as described in Section~\ref{subsec:basic}. A conventional label aggregation algorithm is adopted to infer a $1\times N_i$ resultant initial guess of the ground-truth labels. Similar to the idea explained in Example~\ref{exm:exm1}, the initial guess can then be encoded to a $1\times N_i\times N_c$ slice $\ten{S}$ and concatenated with the raw binarized label tensor. By doing so, we enlarge the size of the label tensor from $N_w\times N_i\times N_c$ to $(N_w+1)\times N_i\times N_c$ and form a target tensor $\ten{T}=[\ten{A};\ten{S}]$, which follows by completion iterations. 
The all-zero $3$-mode fibers of tensor $\ten{A}$ are filled in during the annotation completion step as depicted in~\eqref{eqn:optmization} and~\eqref{eqn:constraint}. At the same time, the imposed low-rank constraint also automatically smooths the noisy label tensor. When the stopping criteria are not satisfied, the bottom annotated slice of the completed tensor is separated and concatenated again with the binarized raw data to form an iterative refinement loop.

\begin{algorithm}[tb]
\caption{Truncated higher-order singular value decomposition (SVD)}
\label{alg:tHOSVD}
\begin{algorithmic}
\STATE {\bfseries Input:} Tensor $\mbox{$\ten{T}\in\mathbb{R}^{I_1\times\cdots\times I_d}$}$, ranks: $R_1, \ldots,R_d$
\STATE {\bfseries Output:} Core tensor \mbox{$\ten{G}\in\mathbb{R}^{R_1\times\cdots\times R_d}$}, factor matrices $\mat{U}^{(1)}, \mat{U}^{(2)},\ldots,\mat{U}^{(d)}$, where $\mat{U}^{(k)}\in\mathbb{R}^{I_k \times R_k}$ for \mbox{$1\leq k\leq d$}.
\FOR{$i=1$ {\bfseries to} $d$}
\STATE $[\mat{L},\mat{\Sigma},\mat{R}^T]\leftarrow$  SVD decomposition of $\mat{T}_{(i)}$
\STATE $\mat{U}^{(i)} \leftarrow$ \text{$R_i$} leading column vectors of $\mat{L}$
\ENDFOR
\STATE $\ten{G}\leftarrow$ $[[\ten{T}; \bm{U}^{(1)T}, \bm{U}^{(2)T},\ldots,\bm{U}^{(d)T}]]$
\end{algorithmic}
\end{algorithm}
\begin{algorithm}[tb]
\caption{Higher-order orthogonal iteration}
\label{alg:HOOI}
\begin{algorithmic}
\STATE {\bfseries Input:} Tensor $\mbox{$\ten{T}\in\mathbb{R}^{I_1\times\cdots\times I_d}$}$, initial rank: $R^0$, ranks: $R_1, \ldots,R_d$.
\STATE {\bfseries Output:} Core tensor \mbox{$\ten{G}\in\mathbb{R}^{R_1\times\cdots\times R_d}$}, factor matrices $\mat{U}^{(1)}, \mat{U}^{(2)},\ldots,\mat{U}^{(d)}$, where $\mat{U}^{(k)}\in\mathbb{R}^{I_k \times R_k}$ for \mbox{$1\leq k\leq d$}.
\STATE Initial factor matrices $\mat{U}^{(k)}\in\mathbb{R}^{I_k \times R^0}$ for \mbox{$1\leq k\leq d$} using Algorithm~\ref{alg:tHOSVD} with $\ten{T}$ and $R_1^0=\ldots=R_d^0=R^0$ as the input.
\WHILE {stopping criteria not satisfied }
\FOR{$i=1$ {\bfseries to} $d$}
\STATE $\ten{A}\leftarrow$ $\ten{T}\, {\times_1}\, \bm{U}^{(1)T} \cdots{\times_{i-1}}\, \bm{U}^{(i-1)T}  {\times_{i+1}}\, \bm{U}^{(i+1)T}$\,  $\cdots{\times_{d}}\, \bm{U}^{(d)T}$ 
\STATE $[\mat{L},\mat{\Sigma},\mat{R}^T]\leftarrow$  SVD decomposition of $\mat{A}_{(i)}$
\STATE $\mat{U}^{(i)} \leftarrow$ \text{$R_i$} leading column vectors of $\mat{L}$
\ENDFOR
\ENDWHILE
\STATE $\ten{G}\leftarrow$ $[[\ten{T}; \bm{U}^{(1)T}, \bm{U}^{(2)T},\ldots,\bm{U}^{(d)T}]]$
\end{algorithmic}
\end{algorithm}

\subsection{Algorithms}
\label{subsec:algo}
Before proposing the pseudocode of the MiSC algorithms, we will briefly introduce the truncated higher-order SVD (HOSVD) procedure~\cite{de2000multilinear} (denoted as Algorithm~\ref{alg:tHOSVD}) and the higher-order orthogonal iteration (HOOI) algorithm~\cite{de2000best} (denoted as Algorithm~\ref{alg:HOOI}), which serve as two cornerstones for the low-rank Tucker completion. 

The ultimate goal of Algorithms~\ref{alg:tHOSVD} and~\ref{alg:HOOI} is to find a (nearly) optimal low-rank Tucker decomposition approximation of the target tensor $\ten{T}$. While finding an exact Tucker decomposition of the multilinear ranks is rather straightforward by directly employing HOSVD algorithm, the Tucker decomposition of ranks smaller than the multilinear ranks found by truncated HOSVD is demonstrated to be non-optimal in~\cite{tensorreview}, in terms of the norm of the difference. Hence we only use a truncated HOSVD step as an initialization procedure. Algorithm~\ref{alg:tHOSVD} takes in a tensor $\ten{T}$ and prescribed Tucker ranks as inputs, and followed by consecutive SVDs operating on different mode matricizations. Factor matrices of the Tucker model are determined by the leading left-singular vectors within each SVD step. Lastly, the core tensor $\ten{G}$ is computed through the full multilinear product of the tensor~\mbox{$\ten{T}\in\mathbb{R}^{I_1\times\cdots \times I_d}$} with matrices $\bm{U}^{(1)T}, \bm{U}^{(2)T},\ldots,\bm{U}^{(d)T}$. An initial Tucker decomposition is thereby obtained. It is worth noting that in the HOSVD algorithm, the mode matricizations are always computed directly from the input tensor. Hence the computation of each of the factor matrices is completely independent of other factor matrices. 

The higher-order orthogonal iteration (HOOI) algorithm was proposed in~\cite{de2000best} to give a low-rank Tucker decomposition with a smaller norm of the difference, without guarantees to converge to the global optimum. In Algorithm~\ref{alg:HOOI}, a Tucker decomposition of prescribed initial ranks is firstly constructed using Algorithm~\ref{alg:tHOSVD}. Then the updates follow an alternating linear scheme (ALS): the factor matrices are updated iteratively by updating one at a time and keeping all the others fixed. In contrast to HOSVD, the update of a factor matrix in HOOI is not independent of others. Specifically, the $i$-mode matricization $\mat{A}_{(i)}$ in HOOI is computed from tensor $\ten{A}$, where \mbox{$\ten{A}\in\mathbb{R}^{R_1\times\cdots\times  R_{i-1}\times I_i\times R^0\times\cdots\times R^0}$} in the first ALS sweep (a full for-loop corresponds to one sweep). Starting from the second ALS sweep,  the size of tensor $\ten{A}$ before $i$-mode matricization becomes \mbox{$R_1\times\cdots\times  R_{i-1}\times I_i\times R_{i+1}\times\cdots\times R_d$}. One can update the Tucker decomposition for a fixed amount of sweeps or until the residual stops decreasing.
\begin{algorithm}[tb]
\caption{Mixed Strategies Crowdsourcing (MiSC)}
\label{alg:MiSC}
\begin{algorithmic}
\STATE {\bfseries Input:} Label matrix $\mbox{$\mat{A}\in\mathbb{R}^{N_w\times N_i}$}$, prior statistics $\ten{S}$, initial rank: $R^0$, ranks: $R_1,R_2,R_3$.
\STATE {\bfseries Output:} Inferred true labels.
\STATE Find $N_c$ by checking the maximum entry of $\mat{A}$
\STATE Initialize conventional aggregation result $\mat{s}$ from $\mat{A}$
\STATE Construct $\ten{A}$ through the following for-loop:
\FOR{$c=1$ {\bfseries to} $N_c$}
\STATE $\ten{A}(:,:,c)\leftarrow$ $\text{double}(\mat{A}==c)$
\ENDFOR
\WHILE {stopping criteria not satisfied }
\STATE $\ten{S}\leftarrow$ binarize $\mat{s}$ as discussed in Example~\ref{exm:exm1}
\STATE $\ten{T}\leftarrow$ concatenate $\ten{A}$ and $\ten{S}$ 
\STATE \mbox{$[\ten{G},\mat{U}^{(1)},\mat{U}^{(2)},\mat{U}^{(3)}]\leftarrow$ $\text{HOOI}(\ten{T},R^0,R_1,R_2,R_3)$}
\STATE $\ten{L}\leftarrow$ $[[\ten{G}; \bm{U}^{(1)}, \bm{U}^{(2)},\bm{U}^{(3)}]]$
\STATE $\mat{s}\leftarrow$ separate the last slice from $\ten{L}$
\ENDWHILE
\STATE $\hat{\mat{A}}\leftarrow$ choose the indices of the maximum values of all $3$-mode vectors in $\ten{L}$
\STATE Infer the true labels from the completed label matrix $\hat{\mat{A}}$ by conventional label aggregation techniques
\end{algorithmic}
\end{algorithm}

We summarize the proposed MiSC algorithms with an exemplary Tucker completion case in Algorithm~\ref{alg:MiSC}. The for-loop realizes the label matrix to label tensor conversion process explained in Example~\ref{exm:exm1}. Within the while-loop, HOOI subroutines are employed to fill in and augment the label tensor. 
One can iterate over the filling process for a prescribed amount of sweeps or until the last slice stops evolving.
After the loops, we binarize the completed 3-way tensor which then follows by a deduction step. The binarization is realized by choosing the indices of the maximum values in each of the $3$-mode vectors. A new $(N_w+1)\times N_i$ label matrix $\hat{\mat{A}}$ is thereby constructed. The most computationally expensive steps in Algorithm~\ref{alg:MiSC} are SVDs in the HOSVD and HOOI subroutines, which have computational complexities of approximately $O((I_1I_2I_3)^2/I_i)$ and $O(I_i(R^0)^4)$ flops, respectively.

\section{Experimental Results}
\label{sec:experiments}
In this section, the proposed mixed complete-aggregate strategies crowdsourcing algorithms are compared with conventional label aggregation methods on six popular datasets, including \textit{Web} dataset~\cite{zhou2012learning}, \textit{BM} dataset~\cite{Mozafari:2014:SUC:2735471.2735474}, \textit{RTE} dataset~\cite{snow2008cheap}, \textit{Dog} dataset~\cite{deng2009imagenet,zhou2012learning}, \textit{Temp} dataset~\cite{snow2008cheap}, and \textit{Bluebirds} dataset~\cite{welinder2010multidimensional}. 
Details of their nonzero rates defined by $\frac{\text{\#worker~labels}}{\text{\#items}\times \text{\#workers}}$, and annotation error rates defined by $\frac{\text{\#wrong~labels}}{\text{\#worker~labels}}$ are recorded in Table~\ref{tbl:2-Dfullcompare}. 

Five conventional crowdsourcing methods are considered in the aggregation step of the proposed MiSC. They are: Majority Voting (MV), Dawid-Skene model + Expectation Maximization
 (DS-EM), Dawid-Skene model + mean field (DS-MF)~\cite{liu2012variational}, and Categorical/ Ordinal Minimax Conditional Entropy ($\text{MMCE}_{\text{(C)}}$, $\text{MMCE}_{\text{(O)}}$). 

We consider three tensor completion algorithms in the MiSC algorithms: LRTC\footnote{\url{http://www.cs.rochester.edu/u/jliu/code/TensorCompletion.zip}}, TenALS\footnote{\url{http://web.engr.illinois.edu/~swoh/software/optspace}}, and Tucker. These methods represent three different approaches towards the completion problem. LRTC~\cite{liu2013tensor} aims at minimizing the sum of nuclear norms of the unfolded matrices. While TenALS~\cite{Prateek2014Provable} and Tucker utilize CP decomposition and Tucker factorization, respectively.

\begin{table*}[ht]
\scriptsize
\centering
\caption{Estimation errors (\%) of pure and mixed strategies on Web, BM, RTE, Gog, Temp, and Bluebirds datasets. Nonzero rates and annotation error rates of datasets are given after their names ($\cdot\%/ \cdot\%$). As an example, the lowest estimation error in the Web dataset comes from the low-rank Tucker completion + $\text{MMCE}_{\text{(O)}}$ aggregation strategies.}
\begin{tabular}{cccccc||cccccc}
\noalign{\hrule height 0.75pt}
\textbf{Web} (3.3/ 63.4) & MV & DS-EM & DS-MF & $\text{MMCE}_{\text{(C)}}$ & $\text{MMCE}_{\text{(O)}}$ & \textbf{BM} (6.0/ 31.1) & MV & DS-EM & DS-MF & $\text{MMCE}_{\text{(C)}}$ & $\text{MMCE}_{\text{O)}}$\\ 
\hline
pure & $ 26.93 $ & $ 16.92 $ & $ 16.10 $ & $ 11.12 $ & $ 10.33 $ & pure & $ 30.4 $ & $ 27.60 $ & $ 26.90 $ & $ 27.10 $ & $ 27.10 $  \\ 
LRTC & $ 26.76 $ & $ 16.55 $ & $16.09$ & $ 11.12 $ & $ 10.33 $ & LRTC & $ 29.25 $ & $27.60  $ & $ 26.90$ & $27.10$ & $27.10$ \\ 
TenALS & $ 26.93 $ & $ 16.77  $ & $ 15.83 $ & $ 11.12 $ & $ 10.33 $ & TenALS & $ 27.60 $ & $ 27.60 $ & $26.90$& $27.10$ & $27.10$ \\ 
Tucker & $10.87$ & $5.77$ & $5.73$ & $6.97$ & $\mathbf{5.24}$ & Tucker & $ 26.50 $ & $27.00$ & $\mathbf{26.20}$ & $ 26.40 $ & $ 26.40 $ \\ 
\hline
\textbf{RTE} (6.1/ 16.3) & MV & DS-EM & DS-MF & $\text{MMCE}_{\text{(C)}}$ & $\text{MMCE}_{\text{(O)}}$ & \textbf{Dog} (9.2/ 30.0) &  MV & DS-EM & DS-MF & $\text{MMCE}_{\text{(C)}}$ & $\text{MMCE}_{\text{(O)}}$\\ 
\hline
pure & $ 10.31 $ & $ 7.25 $ & $ 7.13 $ & $ 7.50 $ & $ 7.50 $ & pure & $ 17.78 $ & $ 15.86 $ & $ 15.61 $ & $ 16.23 $ & $ 16.73 $ \\ 
LRTC & $ 9.25 $ & $7.25  $ & $ 7.00$ & $7.50 $ & $ 7.50 $ & LRTC & $ 15.61 $ & $ 15.61 $ & $ 15.61 $ & $ 15.61 $ & $ 15.61 $ \\ 
TenALS & $ 10.25 $ & $ 7.25 $ & $7.13$ & $7.50 $ & $ 7.50 $ & TenALS & $ 15.86 $ & $ 15.74 $ & $ 15.61 $ & $ 15.86 $ & $ 15.86 $ \\ 
Tucker & $ 8.38 $ & $6.88$ & $\mathbf{6.75}$ & $ 7.50 $ & $7.50 $ & Tucker & $ 15.61 $ & $15.49$ & $\mathbf{15.37}$ & $ 15.86 $ & $ 15.86 $\\ 
\hline
\textbf{Temp} (13.2/ 15.9) & MV & DS-EM & DS-MF & $\text{MMCE}_{\text{(C)}}$ & $\text{MMCE}_{\text{(O)}}$ & \textbf{Bluebirds} (100.0/ 36.4) & MV & DS-EM & DS-MF & $\text{MMCE}_{\text{(C)}}$ & $\text{MMCE}_{\text{(O)}}$\\ 
\hline
pure & $ 6.39 $ & $ 5.84 $ & $ 5.84 $ & $ 5.63 $ & $ 5.63 $ & pure & $ 24.07 $ & $ 10.19 $ & $ 10.19 $& $ 8.33 $ & $ 8.33 $   \\ 
LRTC & $ 5.19 $ & $ 5.63 $ & $ 5.63 $ & $ 5.63 $ & $ 5.63 $ & LRTC & $ 20.37 $ & $ 9.26 $ & $ 9.26 $ & $ 6.48 $ & $ 6.48 $ \\ 
TenALS & $ 5.41 $ & $ 5.63 $ & $ 5.84 $ & $ 5.63 $ & $ 5.63 $ & TenALS & $ 23.15 $ & $ 9.26 $ & $ 9.26 $ & $ 6.48 $ & $ 6.48 $ \\ 
Tucker & $ 5.19 $ & $\mathbf{4.98}$ & $\mathbf{4.98}$ & $ 5.41 $ & $ 5.41 $ & Tucker & $ 19.91 $ & $8.33$ & $9.26$ & $\mathbf{4.63}$ & $\mathbf{4.63}$ \\ 
\noalign{\hrule height 0.75pt}
\end{tabular}
\label{tbl:2-Dfullcompare}
\end{table*}

\subsection{Comparison with State-of-the-Arts}\label{subsec:comparison}
The estimation errors of all five pure conventional label aggregation algorithms and fifteen MiSC approaches on six real-life datasets are reported in Table~\ref{tbl:2-Dfullcompare}. It is shown that the proposed MiSC algorithms achieve lower estimation errors than traditional pure crowdsourcing methods on \emph{all} six datasets. We observe the greatest breakthrough in the Web dataset, where the state-of-the-art estimation errors of around $10.33\%$ are brought down to around $5.24\%$. The second noticeable refinement is in the Bluebirds dataset, where MiSC algorithms produce error rates lower than $5\%$, in contrast to the $>8\%$ state-of-the-art records. For the BM, RTE, Dog datasets, the MiSC algorithm via DS-MF+Tucker also reaches the lowest errors of $26.2\%$, $6.75\%$, and $15.37\%$, respectively, among others.  

\subsection{Pure vs. Mixed Strategies Crowdsourcing}
Besides evaluating all MiSC algorithms together with pure label aggregation methods, we also zoom in on pairwise comparisons between pure and mixed strategies that use the same label deduction approaches. Specifically, in the Web dataset, the MiSC algorithms stacked by DS-EM/ DS-MF and Tucker completion have estimation errors of $5.77\%/ 5.73\%$, compared with the initial $16.92\%/ 16.10\%$ by pure DS-EM/ DS-MF. In the Temp dataset, both DS-EM+Tucker and DS-MF+Tucker improve the performance of their corresponding pure counterparts by approximately one point. In the Bluebirds dataset, mixing MMCE aggregations with Tucker completion helps the crowdsourcing reach the lowest error rate of $4.63\%$ compared with the original $8.33\%$.

Consequently, we have empirically verified that the proposed MiSC algorithms can consistently improve the performance on top of conventional label aggregation schemes. By and large, we also remark that this complete-aggregate two-step looping structure is readily compatible with other state-of-the-art label deduction and tensor completion algorithms.

\subsection{MiSC for Sparse and Noisy Annotations}
\begin{figure}[t] 
\vspace{-0.5em}
\begin{center} 
\includegraphics[width=0.48\textwidth]{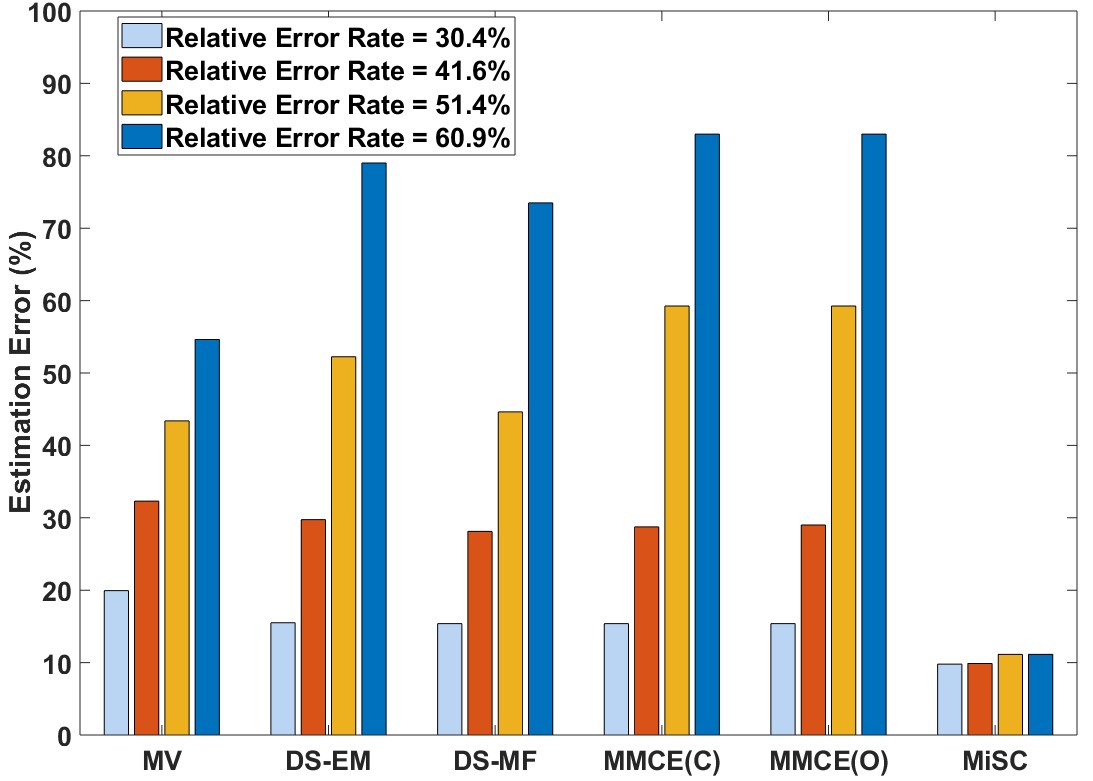}
\caption{Estimation errors ($\%$) of pure and mixed strategies on highly sparse and severely noisy annotations in the RTE dataset.}
\label{fig:high_sparsity}
\end{center}
\vspace{-0.5em}
\end{figure}
As noticed in Section~\ref{subsec:comparison}, although MiSC successfully improves the annotation quality in all six datasets, the striking refinement obtained in Web dataset stands out and raises the question: when will MiSC be remarkably advantageous? Referencing Table~\ref{tbl:2-Dfullcompare}, one can see that the Web dataset has the sparsest and ``poorest'' annotations. Only $88$ items out of $2665$ are labeled on average per worker, and $63.4\%$ of the total $15567$ labels are misleading. We emulate similar scenarios by simultaneously eliminating worker labels from and adding noise to the RTE dataset. The resultant label matrix has a nonzero rate of $3.7\%$, and four different degrees of annotations error rates: $30.4\%, 41.6\%, 51.4\%, 60.9\%$. Figure~\ref{fig:high_sparsity} records the corresponding error rates. We can then see that MiSC is significantly more robust to high sparsity and severe noise. Replacing reliable annotations by erroneous labels only slightly affects the accuracy of MiSC, while the performance of traditional pure label aggregation degrades rapidly. We highlight this important finding since the number and quality of worker labels have a huge implication on the crowdsourcing cost. To this end, the proposed MiSC approach provides a means to substantially relax the labor and labeling quality requirements while maintaining superior accuracy.

\section{Conclusion}
\label{sec:conclusion}
This paper has introduced the mixed strategies crowdsourcing (MiSC) framework, a versatile complete-aggregate two-step iterative procedure, for crowdsourcing tasks. MiSC is readily compatible with various completion techniques and deduction schemes. By integrating tensor completion procedures, and importantly, tensor decomposition methods, into label aggregation, the proposed methods can largely improve the crowdsourcing performance. By further assuming a low-rank Tucker structure, the mixed low-rank Tucker model with conventional label aggregation approaches are shown to be particularly favorable when the annotations are highly sparse and severely noisy. Experiments have shown that MiSC consistently outperforms state-of-the-art methods in terms of estimation errors.

\section*{Acknowledgment}
This work is partially supported by the General Research Fund (Project 17246416) of the Hong Kong Research Grants Council.

\bibliographystyle{named}
\bibliography{ijcai19}

\begin{thebibliography}{}

\bibitem[\protect\citeauthoryear{Cand{\`e}s and Recht}{2009}]{candes2009exact}
Emmanuel~J Cand{\`e}s and Benjamin Recht.
\newblock {Exact matrix completion via convex optimization}.
\newblock {\em Foundations of Computational mathematics}, 9(6):717, 2009.

\bibitem[\protect\citeauthoryear{Chen}{2015}]{chen2015incoherence}
Yudong Chen.
\newblock {Incoherence-optimal matrix completion}.
\newblock {\em IEEE Transactions on Information Theory}, 61(5):2909--2923,
  2015.

\bibitem[\protect\citeauthoryear{Cichocki \bgroup \em et al.\egroup
  }{2015}]{cichocki2015tensor}
Andrzej Cichocki, Danilo Mandic, Lieven~De Lathauwer, Guoxu Zhou, Qibin Zhao,
  Cesar Caiafa, and Anh-Huy Phan.
\newblock Tensor decompositions for signal processing applications: From
  two-way to multiway component analysis.
\newblock {\em IEEE Signal Processing Magazine}, 32(2):145--163, 2015.

\bibitem[\protect\citeauthoryear{Collins and Cohen}{2012}]{collins2012tensor}
Michael Collins and Shay~B. Cohen.
\newblock {Tensor decomposition for fast parsing with latent-variable PCFGs}.
\newblock In {\em NIPS}, pages 2519--2527, 2012.

\bibitem[\protect\citeauthoryear{Dawid and Skene}{1979}]{dawid1979maximum}
Alexander~P. Dawid and Allan~M. Skene.
\newblock Maximum likelihood estimation of observer error-rates using the em
  algorithm.
\newblock {\em Applied statistics}, pages 20--28, 1979.

\bibitem[\protect\citeauthoryear{De~Lathauwer \bgroup \em et al.\egroup
  }{2000a}]{de2000multilinear}
Lieven De~Lathauwer, Bart De~Moor, and Joos Vandewalle.
\newblock A multilinear singular value decomposition.
\newblock {\em SIAM journal on Matrix Analysis and Applications},
  21(4):1253--1278, 2000.

\bibitem[\protect\citeauthoryear{De~Lathauwer \bgroup \em et al.\egroup
  }{2000b}]{de2000best}
Lieven De~Lathauwer, Bart De~Moor, and Joos Vandewalle.
\newblock On the best rank-1 and rank-(r 1, r 2,..., rn) approximation of
  higher-order tensors.
\newblock {\em SIAM journal on Matrix Analysis and Applications},
  21(4):1324--1342, 2000.

\bibitem[\protect\citeauthoryear{Deng \bgroup \em et al.\egroup
  }{2009}]{deng2009imagenet}
Jia Deng, Wei Dong, Richard Socher, Li-Jia Li, Kai Li, and Li~Fei-Fei.
\newblock Imagenet: A large-scale hierarchical image database.
\newblock In {\em CVPR}, pages 248--255, 2009.

\bibitem[\protect\citeauthoryear{Denton \bgroup \em et al.\egroup
  }{2014}]{denton2014exploiting}
Emily~L. Denton, Wojciech Zaremba, Joan Bruna, Yann LeCun, and Rob Fergus.
\newblock Exploiting linear structure within convolutional networks for
  efficient evaluation.
\newblock In {\em NIPS}, pages 1269--1277, 2014.

\bibitem[\protect\citeauthoryear{Imaizumi \bgroup \em et al.\egroup
  }{2017}]{imaizumi2017tensor}
Masaaki Imaizumi, Takanori Maehara, and Kohei Hayashi.
\newblock On tensor train rank minimization: Statistical efficiency and
  scalable algorithm.
\newblock In {\em NIPS}, pages 3933--3942, 2017.

\bibitem[\protect\citeauthoryear{Jain and Oh}{2014}]{Prateek2014Provable}
Prateek Jain and Sewoong Oh.
\newblock Provable tensor factorization with missing data.
\newblock In {\em NIPS}, pages 1431--1439. Curran Associates, Inc., 2014.

\bibitem[\protect\citeauthoryear{Karger \bgroup \em et al.\egroup
  }{2011}]{karger2011iterative}
David~R. Karger, Sewoong Oh, and Devavrat Shah.
\newblock Iterative learning for reliable crowdsourcing systems.
\newblock In {\em NIPS}, pages 1953--1961, 2011.

\bibitem[\protect\citeauthoryear{Kolda and Bader}{2009}]{tensorreview}
Tamara~G. Kolda and Brett~W. Bader.
\newblock Tensor decompositions and applications.
\newblock {\em SIAM Review}, 51(3):455--500, 2009.

\bibitem[\protect\citeauthoryear{Levin}{1963}]{levin1963three}
Joseph Levin.
\newblock {\em Three-Mode Factor Analysis}.
\newblock PhD thesis, University of Illinois at Urbana-Champaign, 1963.

\bibitem[\protect\citeauthoryear{Li and Jiang}{2018}]{li2018Multi}
Shao-Yuan Li and Yuan Jiang.
\newblock Multi-label crowdsourcing learning with incomplete annotations.
\newblock In {\em PRICAI}, pages 232--245, 2018.

\bibitem[\protect\citeauthoryear{Liu \bgroup \em et al.\egroup
  }{2012}]{liu2012variational}
Qiang Liu, Jian Peng, and Alexander~T. Ihler.
\newblock Variational inference for crowdsourcing.
\newblock In {\em NIPS}, pages 692--700, 2012.

\bibitem[\protect\citeauthoryear{Liu \bgroup \em et al.\egroup
  }{2013}]{liu2013tensor}
Ji~Liu, Przemyslaw Musialski, Peter Wonka, and Jieping Ye.
\newblock {Tensor completion for estimating missing values in visual data}.
\newblock {\em IEEE Transactions on Pattern Analysis and Machine Intelligence},
  35(1):208--220, 2013.

\bibitem[\protect\citeauthoryear{Liu \bgroup \em et al.\egroup
  }{2015}]{liu2015learning}
Pengfei Liu, Xipeng Qiu, and Xuanjing Huang.
\newblock Learning context-sensitive word embeddings with neural tensor
  skip-gram model.
\newblock In {\em IJCAI}, pages 1284--1290, 2015.

\bibitem[\protect\citeauthoryear{Mozafari \bgroup \em et al.\egroup
  }{2014}]{Mozafari:2014:SUC:2735471.2735474}
Barzan Mozafari, Purna Sarkar, Michael Franklin, Michael Jordan, and Samuel
  Madden.
\newblock Scaling up crowd-sourcing to very large datasets: A case for active
  learning.
\newblock {\em PVLDB}, 8(2):125--136, 2014.

\bibitem[\protect\citeauthoryear{Novikov \bgroup \em et al.\egroup
  }{2015}]{novikov2015tensorizing}
Alexander Novikov, Dmitry Podoprikhin, Anton Osokin, and Dmitry Vetrov.
\newblock Tensorizing neural networks.
\newblock In {\em NIPS}, pages 442--450, 2015.

\bibitem[\protect\citeauthoryear{Raykar \bgroup \em et al.\egroup
  }{2009}]{raykar2009supervised}
Vikas~C. Raykar, Shipeng Yu, Linda~H. Zhao, Anna Jerebko, Charles Florin,
  Gerardo~H. Valadez, Luca Bogoni, and Linda Moy.
\newblock Supervised learning from multiple experts: whom to trust when
  everyone lies a bit.
\newblock In {\em ICML}, pages 889--896, 2009.

\bibitem[\protect\citeauthoryear{Signoretto \bgroup \em et al.\egroup
  }{2014}]{signoretto2014learning}
Marco Signoretto, Quoc~Tran Dinh, Lieven De~Lathauwer, and Johan A.~K. Suykens.
\newblock {Learning with tensors: a framework based on convex optimization and
  spectral regularization}.
\newblock {\em Machine Learning}, 94(3):303--351, 2014.

\bibitem[\protect\citeauthoryear{Snow \bgroup \em et al.\egroup
  }{2008}]{snow2008cheap}
Rion Snow, Brendan O'Connor, Daniel Jurafsky, and Andrew~Y. Ng.
\newblock Cheap and fast---but is it good?: evaluating non-expert annotations
  for natural language tasks.
\newblock In {\em EMNLP}, pages 254--263, 2008.

\bibitem[\protect\citeauthoryear{Suzuki}{2015}]{suzuki2015convergence}
Taiji Suzuki.
\newblock {Convergence rate of Bayesian tensor estimator and its minimax
  optimality}.
\newblock In {\em ICML}, pages 1273--1282, 2015.

\bibitem[\protect\citeauthoryear{Tucker}{1963}]{tucker1963implications}
Ledyard~R. Tucker.
\newblock Implications of factor analysis of three-way matrices for measurement
  of change.
\newblock {\em Problems in measuring change}, 15:122--137, 1963.

\bibitem[\protect\citeauthoryear{Welinder \bgroup \em et al.\egroup
  }{2010}]{welinder2010multidimensional}
Peter Welinder, Steve Branson, Pietro Perona, and Serge~J. Belongie.
\newblock The multidimensional wisdom of crowds.
\newblock In {\em NIPS}, pages 2424--2432, 2010.

\bibitem[\protect\citeauthoryear{Whitehill \bgroup \em et al.\egroup
  }{2009}]{whitehill2009whose}
Jacob Whitehill, Ting fan Wu, Jacob Bergsma, Javier~R. Movellan, and Paul~L.
  Ruvolo.
\newblock Whose vote should count more: Optimal integration of labels from
  labelers of unknown expertise.
\newblock In {\em NIPS}, pages 2035--2043, 2009.

\bibitem[\protect\citeauthoryear{Zhang \bgroup \em et al.\egroup
  }{2016}]{zhang2016spectral}
Yuchen Zhang, Xi~Chen, Dengyong Zhou, and Michael~I. Jordan.
\newblock Spectral methods meet em: A provably optimal algorithm for
  crowdsourcing.
\newblock {\em The Journal of Machine Learning Research}, 17(1):3537--3580,
  2016.

\bibitem[\protect\citeauthoryear{Zhao \bgroup \em et al.\egroup
  }{2015}]{zhao2015bayesian}
Q.~Zhao, L.~Zhang, and A.~Cichocki.
\newblock Bayesian {CP} factorization of incomplete tensors with automatic rank
  determination.
\newblock {\em IEEE Transactions on Pattern Analysis and Machine Intelligence},
  37(9):1751--1763, 2015.

\bibitem[\protect\citeauthoryear{Zhou and He}{2016}]{zhou2016crowdsourcing}
Yao Zhou and Jingrui He.
\newblock Crowdsourcing via tensor augmentation and completion.
\newblock In {\em IJCAI}, pages 2435--2441, 2016.

\bibitem[\protect\citeauthoryear{Zhou \bgroup \em et al.\egroup
  }{2012}]{zhou2012learning}
Denny Zhou, Sumit Basu, Yi~Mao, and John~C. Platt.
\newblock Learning from the wisdom of crowds by minimax entropy.
\newblock In {\em NIPS}, pages 2195--2203, 2012.

\end{thebibliography}

\end{document}